Why are there so few women in physics? Reflections on the experiences of two women

Danny Doucette and Chandralekha Singh
Department of Physics and Astronomy, University of Pittsburgh, Pittsburgh, PA 15260

**Abstract**: Some of the reasons for the underrepresentation of women in physics are evident in the reflections of two undergraduate women. Leia is a chemistry major who loves college-level physical chemistry and quantum mechanics but does not identify with the discipline of physics, partly because she has a low level of self-efficacy as a physicist and has received very little recognition for her work and learning in physics. Paulette is a physics major who loves physics but feels isolated by the current physics learning environment. She reluctantly dropped honors introductory lab after being snubbed by her male classmates who partnered with one another, leaving her to work alone. Paulette's experiences with condescending male professors activated a stereotype threat about who can succeed in physics that caused her to disengage in class. We also discuss what these women felt has helped them so far and explore their suggestions for what would help women in physics courses as they pursue their quest for a physical science degree.

**Introduction**

Issues related to why women are underrepresented in physics are being investigated from various angles (e.g., see Refs. [1-6]). While some women experience explicit and implicit bias while studying physics, some may experience being recognized as a person who can excel in physics and mentoring by physics instructors [7-11] differently from men even in the absence of discrimination [12]. Women can also be affected by isolation and stereotype threat [13-15]. Here we present the voices of two white American undergraduate women in their second and third years at a large public university to shed light on experiences in physics that they found alienating. Both were interviewed about their experiences for one hour each using a semi-structured protocol. All names are pseudonyms.

**Leia: Self-Efficacy and Recognition**

Self-efficacy, the belief in one's ability to be successful at a given task [16], has been shown to be an important factor in students' career decisions, enrollment in STEM courses, and persistence toward long-term goals [7, 8]. However, research has shown a sizable gender gap in the self-efficacy of students enrolled in introductory physics courses, in which women consistently have a lower level of self-efficacy than their male peers, even if those men have lower grades [10]. Leia explains what this paradoxical self-efficacy difference looks like, from a student's perspective: "There would be times when I would feel like I am not good at physics, I am not good at it. But we would get tests back… I was very comparable to them but I still felt like, oh, it is not my thing, I am not very good at it". While Leia knew, on some level, that she was doing just as well as other students in her class, at another point in the interview she described hypothetical physics students who were far superior to her: "Physics I and II [lecture] were a little fast for me but a good speed for everyone… but the physics lab seemed a lot slower-paced so it was really good for me, but it was kind of boring for other people that were very, very, very good at physics."

While Leia had a low physics self-efficacy, she was confident about her capabilities in chemistry. She recounted how getting a very good grade on the physical chemistry (P-chem) midterm exam in the current semester was something that boosted her self-efficacy and increased her interest in P-chem to the point where she was almost looking forward to her final exam: "I just got the grade back. I almost cried because this is the best I have done on a chemistry test. And then I just felt so much more confident to take the final. You know, like, I am REALLY good at P-chem… It was really nice, like, YES! All my hard work is paying off finally." Leia clearly has the ability to be successful, and she has a mindset that supports her in this closely-related field to physics. However, the fact that she feels so negative about her physics abilities serves to highlight how threatening an experience the introductory physics class can be for some women, even though all students are ostensibly treated equally.

In order to understand how Leia could have a positive view of physical chemistry while being apprehensive about physics, we asked her about the recognition she has received as a chemistry student. Students who perceive that they are being positively recognized as members of a science community often experience a boost to their identity as scientists [17, 18]. Leia noted that in her introductory physics course she found support from a female undergraduate teaching assistant from the Undergraduates Teaching Undergraduates (UTU) program. She felt that this was a person who was empathetic, and who understood her and her questions. She also felt she could talk to the teaching assistant about physics at her level and she did not feel anxious talking to her. She was so positively impacted by the support she received from the female physics UTU experience as a student that she herself decided to be a UTU in physical chemistry and noted that the recognition from the students she was teaching was the best part of the experience. "I think honestly the best part was the last session. People were excited to leave, it was the last lab, but then they would turn around and [say] 'oh but we would never see you again' … that was the best part. I think I made their experience in the lab like a little bit better." Further discussions suggest that these UTU experiences, both from the perspective of a student (asking a female UTU for help in the physics context) and as a UTU teaching students in the chemistry context, were empowering for her, and that the recognition she received from both types of experiences helped improve her sense of belonging and self-efficacy.

When discussing what we can do to help students like her, Leia talked about the small class size in her physical chemistry course this current term, "My class size got smaller so I felt like the professor knew my name. That was the first time, well in a science class, a professor really knew my name… So it is just that the higher you go, the smaller the classes, the better you get to know the people. That is the best part for me." Leia further added that smaller classes would really help her because the instructor will be able to communicate with her personally saying, "Let's say I did really poorly 'Leia, what happened' or if I did really well 'Great job Leia!' … I like that a whole lot better than huge, huge classes…". She then reminisced about a non-science course saying, "I had a public speaking course which is my favorite course here so far and [the instructor] is phenomenal! She would cross the street to just say hello. So yeah it definitely, definitely made me feel really welcome." It appears that a smaller class helped to improve Leia's sense-of-belonging and self-efficacy by making it more likely that she would experience positive interactions and receive recognition as a scientist from her instructors. This recognition, once internalized, may have reduced the ambiguity she felt in how she was performing relative to others, and may have helped increase her self-efficacy and sense of belonging. This process may be especially relevant for some students who may not have previously had opportunities to learn physics and interact with content area experts, such as students like Leia who attended small or under-resourced high schools.

**Paulette: Isolation, Stereotype Threat, and Mentorship**

Another difficulty faced by women in physics is the challenge of isolation. Being the only member of a visible minority in a cohort, lab, or department can lead a student to being excluded from productive study groups, missing out on informal forms of support, and even having vital materials withheld by peers [5]. Paulette's experiences in the introductory honors lab are an example of this. She was the only woman in the lab course and consistently found herself working alone, even though she was friends with men in the lab. "[The lab] was really, really cool but it's very unguided... I went in a month and was like, I can't do this anymore. I was the only girl in that class, which is… not ideal. It just doesn't make you feel very comfortable, ever." Paulette explained how she ended up working alone by describing how the male students paired up amongst themselves: "[the other students] were like, hey do you want to do the photoelectric effect with me, do you want to do blackbody radiation with me? … I already knew them, and that was still kind of hard." As the semester went on, Paulette's experiences in the lab became frustrating and exhausting: "it's like, you say stuff, and people don't really listen to you. It's not necessarily intentional. But you suggest something, and people just ignore you. So, you have to be a bit more forceful, and sometimes you just don't have the energy for that… after a certain point I didn't want to go to that class anymore, and it was, I'm just going to withdraw. I didn't want to deal with the mental stress of that. Or, you know, pay money for a class I didn't want to go to".

Paulette also faced the challenge of stereotype threat as a woman in a male-dominated field. Stereotype threat refers to fear of confirming a negative stereotype about oneself because of one's association with a stereotyped group (e.g., women in physics). Stereotype threat can increase anxiety and rob students of the limited cognitive resources and thus lead to deteriorated performance. Stereotype threat has been posited as a possible explanation for the underperformance of traditionally marginalized groups in physics [14, 15, 19, 20] and Paulette's experiences provide a vivid example. Paulette began by articulating a stereotype that impacted her success in class: fear that her instructors were judging her for being a woman and being condescending in their responses to her questions. She mentioned that some of her male physics professors "tend to have an air where it's like, this is the naïve solution, this is so trivial... Sometimes there's condescension, which makes you not want to ask questions anymore." Consequently, she feared raising her hand to ask questions in class: "It can be especially hard when you have that question in the back of your mind which is like, are they being condescending because they think I can't do it, or because I'm a woman?" This fear and response became a threat and the condescension had a negative impact on Paulette's sense of belonging. She explained how she, and her female classmates, experienced a lower sense of belonging than their male peers, and how this impacted them differently: "I feel like most of the physics majors at one point are like, I'm too dumb, everyone else is smarter than me. But I've only ever heard the women contemplating… I don't belong in this and maybe I should drop the major. I've never heard any of the guys say that. And I don't know… there's only one thing that can really stem from!" Paulette described this gender difference as arising from feelings of isolation, gendered microaggressions, and the ubiquity of stereotypes such as the belief that physics is a field for `brilliant' men. We can imagine how the same process may have hurt Paulette in other circumstances by noting that, since human working memory has a limited capacity while problem solving [21], lower self-efficacy and the resulting anxiety associated with stereotypes could have robbed Leia of some of her cognitive resources while learning physics or taking physics quizzes and exams, and she could potentially have done even better than she actually did if she did not have such concerns.

Paulette also described positive experiences during our interview, such as how she feels empowered with the support of a female faculty mentor. Mentoring can play an essential role for students, providing essential relational resources to support the development of students' sense of belonging and identity as scientists [11, 22]. Similarly, role models can help to counter stereotypes about who can succeed physics. In her interview, Paulette noted that both faculty mentors and peers can serve to provide support and reassurance. Paulette lamented not having a female professor in her early physics classes: "I think it would be nice to have a female professor... Because I [would] just feel more comfortable talking to her. Until then I've just had men." When asked what kinds of things have supported her in college so far, Paulette described her current research internship with a female professor, "I have my female mentorship thing with [a professor of physical chemistry], which was definitely on purpose. I looked for female PIs." She was very happy about her relationship with her mentor and noted, "We meet every week. She's like, how are you doing, where are you at? … The time I know I have problems, I'll talk to her then, she'll talk me through it, and I can keep doing my work." Support can also come from peers. Reminiscing on the lab Paulette said, "I know, almost for a fact, that if my friend Amy had been in that class I would have stayed in that class because we would have partnered on pretty much everything." Paulette also found comfort in talking to a female physics teaching assistant, "My first semester, the TA, she's a graduate student here. That was nice, I just automatically felt more comfortable asking questions."

**Implications and Summary**

Reflections on the interviews with Leia and Paulette lead to the following suggestions for instructors:
1. Ensure that students are not isolated. In settings where group work is expected, or necessary, ensure that students partner equitably even if it means being "uncool". As Paulette suggests, "making the teachers assign partners, as uncool as that is, would definitely help with that kind of thing. Because then you don't feel like you're forcing someone to be your partner." Just as research suggests that instructors be careful to avoid isolating underrepresented minority students in majority-dominated groups [23], we suggest that instructors should also be careful to avoid isolating underrepresented minority students into doing solo work because they cannot always find a partner.
2. Make sure to recognize students for their successes, including the often-overlooked success of simply making progress in their studies. Also, take interest in students' well-being. As Leia describes, simply acknowledging students inside or outside of class can have a long-lasting positive impact. For instructors, this could mean learning students' names and using them, acknowledging students when you see them outside of the classroom, and providing friendly encouragement when students do well on an assignment.
3. Take responsibility for the social dynamics in your classroom and pay attention to students who may be feeling under-supported. As Paulette explains, "I think it was probably pretty obvious that I wasn't happy in the class, and neither of the two professors said anything to me about it, even though it was only 10 [other] people." This might include establishing and reinforcing positive norms for community work and engaging in frequent check-ins with student groups.
4. Where possible, advocate for smaller classes. As Leia pointed out, smaller classes and recitations give instructors and peers the opportunity to get to know each other better. Smaller classes may make it easier to attend to students' self-efficacy, pay attention to social dynamics in the classroom and ensure everyone has positive experiences. Similarly, consider establishing or supporting study groups and affinity groups in your department. Paulette proposes: "Women study groups? Bring your homework and someone who's taken the class before will help you. I think for physics that would definitely be helpful. Or at least, like, a networking kind of thing? It'd be nice to know that there's a woman that has done it before."

These interviews shed light on the experiences of two undergraduate female students majoring in chemistry and physics to give a glimpse of how societal stereotypes can activate stereotype threats [19-21] and how, without adequate support and even in the absence of discrimination, women in physics courses can develop low levels of self-efficacy and be made to feel isolated. As educators, it is critical to reflect upon the experiences of these women since they can impact students' classroom experiences, grades, persistence, and career choices. Physics teachers can benefit from reflecting on these students' narratives in order to create equitable learning environments in which all students can thrive and excel.